\documentclass[aip,amsmath,amssymb,reprint,]{revtex4-1}

\usepackage{graphicx}

\begin{document}

\title{Bose-Einstein Condensation of Magnons in
       NiCl$_{2}$-4SC(NH$_{2}$)$_{2}$}

\author{Armando Paduan-Filho}

\affiliation{Instituto de F\'{i}sica, Universidade de S\~{a}o
Paulo, 05314-970 SP, Brazil\\}

\date{\today}

\begin{abstract}

A Bose-Einstein condensation (BEC) has been observed in magnetic
  insulators in the last decade. The bosons that condensed are
  magnons, associated with an ordered magnetic phase induced by
  a magnetic field. We review the experiments in the
  spin-gap compound NiCl$_{2}$-4SC(NH$_{2}$)$_{2}$, in which the
  formation of BEC occurs by applying a magnetic field at
  low temperatures. This is a contribution to the celebration for the
  50th anniversary of the Solid State and Low Temperature Laboratory
  of the University of S\~{a}o Paulo, where this compound was first magnetically characterized.

\end{abstract}
 \maketitle

\section{I. INTRODUCTION}
Macroscopic systems governed by quantum mechanics of interacting
particles attract a great deal of interest. Cold atoms and quantum
magnets, whose total spin is an integer, have interesting
similarities that show the common physics of these two seemingly
different realizations \cite{1,2,3,4,5,6,7,8,9,10,11}. All atoms
with an even number of neutrons satisfy Bose statistics, which
accounts for about 75 per cent of the atoms in the periodic table.
Based on the Bose-Einstein statistics a gas of non-interacting
massive bosons condenses below a certain temperature $T_{BEC}$, in
which the Bose-Einstein condensation (BEC) occurs. This is a
macroscopic quantum phenomenon characterized by spontaneous
quantum coherence persisting over macroscopic length and time
scales. In dilute atomic gases this phenomenon was realized
experimentally for cold atoms. Several quantum spin systems in
solids, which show a magnetic-field induced transition, are
expected to also show condensation above or below a certain
critical field \cite{2,7,12}. Studies have shown that the magnetic
system can be mapped non-locally onto a set of weakly interacting
bosons on a lattice.

The sorting phase can be described as a BEC of bosonic
quasi-particles, in which the magnetic field acts to preserve the
number of bosons. Therefore, the tuning parameter to induce
condensation in spin-ordered systems is not the temperature, but
the magnetic field. For particles as well as magnons, a
macroscopic number of bosons condense into a single quantum
state\textemdash the state of lowest energy. The quantum coherence
of Bose-Einstein condensation dates back to the prediction of
Einstein, based on Bose's work, in 1924.

In the diluted BEC the macroscopic wave function is directly
connected with the microscopic energy levels, providing a complete
description of these phenomena in terms of the Gross-Pitaevskii
equation. The concept of a coherent macroscopic matter wave in
interacting many-body systems is independent of a detailed
microscopic understanding of particles. The intricacies of the
many-body problem with interactions that lead to non-separable
Hamiltonians are solved by this equation, which introduces
effective potentials that are simpler than the original
interactions, which in turn renders the physical problem more
tractable.

Experimental evidence of the BEC in confined weakly interacting
gases was produced by E.~A.~Cornell, W.~Ketterle, and C.~E.~Wieman
in 1995, leading to a Nobel Prize in 2001. That BEC would occur in
quantum magnets was first predicted in 1991, and several reports
describing real systems were published after 2000 \cite{7,13,14}.
The possibility of experimental investigation of the BEC in
quantum magnets has led to a deeper understanding of the ground
states in strongly correlated systems.  Recent studies concerning
the BEC of bosons in magnetic systems ranged from a 3$D$ weakly
coupled $S$=1/2 dimer compound, $\rm TlCuCl_{3}$ \cite{7} (the
first system in which an identification was attempted with BEC), a
quasi-2$D$ spin system, $\rm BaCuSi_{2}O_{6}$ \cite{13}, to a
quasi-1$D$ $S$=1 chain with single-ion anisotropy,
NiCl$_{2}$-4SC(NH$_{2}$)$_{2}$ (DTN) \cite{14}. In addition to
these three compounds, other gapped spin systems share the same
physical picture of the BEC, most of which comprise Cu dimers.

The focus of this article is to review systematic concepts and
experimental realizations of a transition to a BEC phase in DTN,
where bosons are quanta of magnetic excitations in a magnetically
ordered ensemble of magnetic moments. Although there is no
consensus on a specific name for the magnetic excitations, we will
use the word magnon for the excitations in insulating systems
generated by the application of a magnetic field. This
denomination follows the nomenclature used by the first papers on
BEC in magnetic insulators, since elementary excitations in
antiferromagnetic systems are called magnons \cite{2,7}.

We hope that the experimental evidence for BEC in DTN will
stimulate new research in this fascinating quantum phenomenon.

The outline of this paper is as follows. Section 2 reviews the
concepts of the ideal BEC at finite temperature and the nature of
BEC when magnons are induced by the application of a magnetic
field. Section 3 presents the magnetic properties of DTN,
introducing the Hartree-Fock-Popov analysis of the hard-core boson
Hamiltonian. Section 4 presents experimental works that lead to
the interpretation of the induced phase diagram as a BEC. Section
5 summarizes the results obtained for DTN in the framework of the
BEC of magnons.

\begin{figure}
\begin{center}
\includegraphics[width=7cm,keepaspectratio=true]{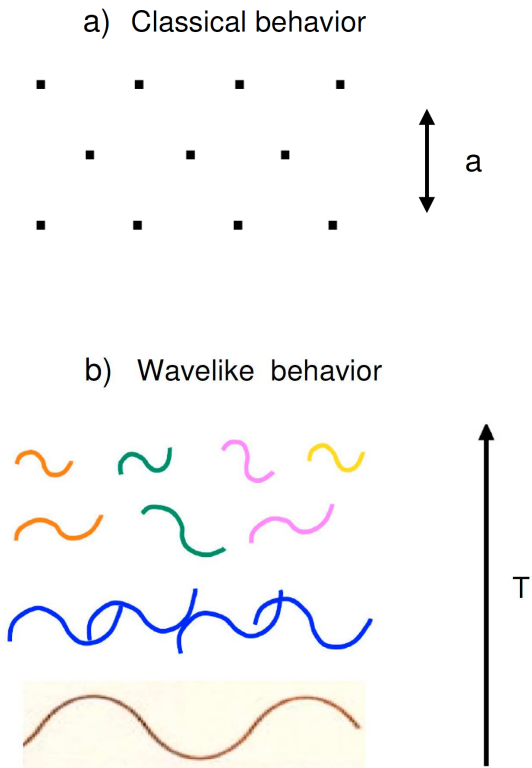}

\caption{\label{fig1} (a) Classical representation of a solid. $a$
is
  the average distance between the particles. (b) In the wavelike
  representation the position of the particles is given by the thermal
  $de$ $Broglie$ wavelength $\lambda$. Decreasing the temperatures,
  the wave-packet becomes correlated for $\lambda$ $\approx$ a. For
  particles in the same quantum state, such as bosons at very low
  temperatures, the system is represented by a single macroscopic
  wave-function.}
\end{center}
\end{figure}

\section{II. BOSE-EINSTEIN CONDENSATION}

According to the Heisenberg uncertainly principle the position of
a particle is smeared out over a distance given by the thermal
$de$ $Broglie$ wavelength $\lambda = h/(2\pi mk_{B}T)^{1/2}$ ,
where $\ k_{B}$ is the Boltzmann constant, $m$ is the particle
mass, and $T$ is the system temperature \cite{5,11}. At room
temperature $\lambda$ is typically one hundred thousand times
smaller than the average spacing between the particles, $a$. This
means that the matter waves of the individual particles are
uncorrelated, and the system can therefore be described by
classical Boltzmann statistics. As the system is cooled,
eventually the distance between two particles becomes the same as
the $de$ $Broglie$ wavelength, $a$ $\approx$ $\lambda$
\cite{1,11}. The packet-functions of adjacent particles overlap,
causing the atoms to lose their identity, and the behavior of the
system becomes strongly correlated. The system is now governed by
quantum statistics. The density distribution of the condensate is
represented by a single macroscopic wave function with a
well-defined amplitude and phase, which are the order parameters
of the system, and a Bose Einstein condensate is formed. Figure~1
shows illustrates this.

The Bose-Einstein condensation is associated with bosonic
particles at low temperatures that condensate into a quantum state
at a temperature $T_{BEC}$. At low temperatures, the density of
bosons with mass $m$ may be expressed by the power law dependence
equation within the BEC picture as

\begin{equation}
  \rho=\zeta(3/2)\left(\frac{mk_{B}}{2\pi\hbar^{2}}\right)^{3/2}
  T_{BEC}^{3/2},
\end{equation}
where $\zeta$ is the Riemann zeta function, and $\zeta(3/2) =
2.612$ \cite{1,7,9}.

For diluted bosons with small interaction, meaning that the
two-particle scattering length $x$ is much shorter than the
interatomic distance, the observed $T_{BEC}$ agrees well with the
theoretical expectation for free particles. For less diluted
systems, the interaction between bosons gives rise to a small
shift of $\ T_{BEC}$ \cite{4,5}:
\begin{equation}
\Delta(T_{BEC})/T_{BEC} = c \rho^{1/3}x.
\end{equation}

Various authors have attempted to determine the value of $c$,
which was found to be in the range $1-2$, although all authors
agree on the functional form.

\subsection{2.1. Candidates for BEC}

There are several candidates for BEC. A first attempt to
experimentally observe BEC was made with superfluidity. When
$^{4}$He is cooled to a critical temperature of 2.17\,K, the
liquid density drops, and fractions of the liquid become
superfluid. The strong interaction of $^{4}$He in liquid form
prevents perfect condensation. The formation of Cooper pairs of
electrons, which act like bosons, is also evidence of the
condensation effect, which produces superconductivity \cite{1,15}.
However, in these systems, a macroscopic wave function provides
only a phenomenological description of the superfluid state.

By contrast, the macroscopic wave function of pure condensates is
directly connected with the microscopic degrees of freedom,
providing a complete and quantitative description of both static
and dynamic phenomena. This description should cover systems
showing a true BEC phase. Some systems were cited, such as
excitons-polaritons, which are light-mass quasi-particles usually
produced by optical excitations in semiconductors, and photons
confined in a cavity \cite{1,11,15,16,17}. However, excitons have
a very short lifetime and the photons are absorbed by the walls.
Because the BEC requires a thermalization time to condense in
which the number of particles is conserved, their characteristics
almost prevent these systems from becoming BECs. All these systems
are believed to be defective condensates.

Recently it has been reported that a gas of magnons at room
temperature can be continuously overpopulating the lowest energy
level, which shows characteristics of BEC, when it is driven by
microwave pumping.\cite{18,19}

On the other hand, diluted magnons in equilibrium can be pure
condensates at low temperatures, just like cold gas.

\subsection{2.2. BEC of Magnons}

Magnetic insulators with magnetic ions are characterized by a gap
separating the singlet ground state from the lowest-energy excited
state. If the gap is closed by the Zeeman effect, the resulting
quasiparticles can undergo the Bose-Einstein condensation
transition.

The BEC phase in quantum magnets is an XY antiferromagnetic state
in which the spins spontaneously choose a particular orientation
in the XY plane. The number of bosons is proportional to the net
magnetization, i.~e., the $S^{z}$ component of the spins. On the
other hand, the number of condensed bosons is proportional to the
antiferromagnetically ordered $S^{x}$ component of the spins. Thus
the XY antiferromagnetic order parameter maps directly onto the
order parameter for Bose-Einstein condensation. It has a magnitude
(size of the spins in the spin language, or number of bosons in
the boson language) as well as a phase (direction of the spins).
In the particle language, the boson number can be tuned by the
magnetic field. For this to occur, the system must be an XY
magnet, i.~e., must have uniaxial symmetry in the plane
perpendicular to the applied magnetic field. Therefore only
certain quantum magnets can be treated as BECs. We note that in
real systems there will always be terms that break the uniaxial
symmetry, such as dipole-dipole interactions and spin-orbit
interactions. Consequently the theory is valid only when these
terms are significantly smaller than the temperature. At very low
temperatures any real quantum magnet will cross over to the Ising
universality class.

Bosons can be achieved by creating and maintaining a large number
of quasi-particles in a system. One special case is the creation
of magnons at the induced magnetic-ordering phase. Figure~2 shows
the phase diagram in the temperature-versus field plane of an
induced antiferromagnetic ordering in a field whose magnetic ions
are strongly correlated. Magnons can be created and, under special
conditions, condensed into the ordered phase \cite{2,8,14}.

A procedure for studying the properties of the phase diagram is to
understand the transitions that appear when the boundaries
separating the phases are crossed \cite{7,20,21,22}. When a border
is crossed vertically to enter the ordered phase a thermal phase
transition results, driven by thermal fluctuations. The magnetic
and thermal energies compete. At temperatures near zero, by
contrast, there are no thermal fluctuations and a quantum
transition results, driven by quantum fluctuations. The magnetic
energy and the zero-point energies compete. This second-order
quantum transition to the BEC phase is free of interactions.
However, including even week interactions into the description of
the condensation has proven to be a great challenge, which was
overcome by introducting effective interactions \cite{7,10,20,21}.
These interactions, well-approximated by a contact
pseudo-potential, are central to arriving at the Gross-Pitaevskii
equation. Understanding the behavior of the phase transition near
zero temperature is understanding the quantum behavior of the
system.  The transition to the BEC state is characterized by
specific characteristics of the thermodynamic properties.

We studied BEC in magnons because condensation occurs at
temperatures that are much more accessible than the condensation
of particles.

\begin{figure}
\begin{center}
\includegraphics[width=7cm,keepaspectratio=true]{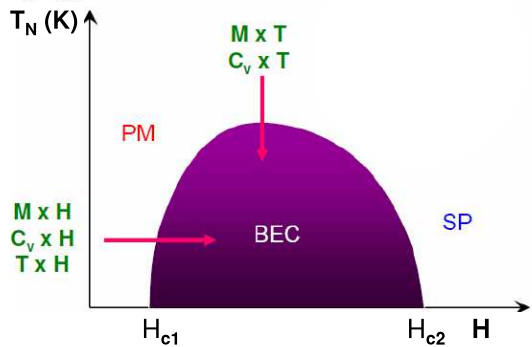}
\caption{\label{fig2} The phase diagram from the normal to the BEC
  state can be studied via thermodynamic properties such as
  magnetization, temperature and specific heat, which are measured
  when the phase boundary is crossed by sweeping the field or the
  temperature, as indicated by the arrows.}
\end{center}
\end{figure}

Since according to~(1) $T_{BEC}$ is proportional to
$\rho^{2/3}/m$, we would like to maximize $\rho$ to observe the
BEC at easily accessible temperatures. However, that equation
assumes that the bosons are free particles and is therefore only
valid in the limit of low density. We can overcome this problem by
considering that $T_{BEC}$ is also dependent on the boson mass
$m$. By minimizing $m$, we can observe the BEC at relatively high
temperatures without sacrificing the condition of low boson
density. In quantum magnets, the boson mass is sufficiently small,
so that the BEC can be realized at temperatures of a few Kelvin,
in contrast to cold atoms, in which temperatures in the
nano-Kelvin range are needed to observe the BEC in magnetic
optical traps \cite{19,21}.

\section{III. MAGNETIC PROPERTIES OF DTN-BEC}

In addition to the properties of NiCl$_{2}$-4SC(NH$_{2}$)$_{2}$,
one needs well-prepared samples for studies under laboratory
conditions. Single crystals of dichloro-$tetrakis$thiourea-nickel
(DTN) were grown by dissolving a large excess of nickel chloride
in a saturated solution of thiourea \cite{23}. The crystals grow
after about two weeks in a solution maintained at a temperature of
35 degrees centigrade. Yellow-brown single crystals are tetragonal
prisms along the $c$-axis (Fig. 3a). The space group of DTN is
$I4$, with two molecules in the unit cell, whose dimensions are
$a=b=9.558$ \AA $ $ and $c=8.981$ \AA. The $S = 1$ Ni atoms form
two interpenetrating tetragonal lattices. The thioureas are
bounded to the nickel by means of the sulfur atoms in a square
array. The spin must be in an environment of approximate uniaxial
symmetry (Fig. 3b). The metal is in a noncentrosymmetric
environment of fourfold symmetry. All molecules are oriented in
the same way and held together by Cl-N hydrogen bonds. The
Cl-Ni-Cl axes all lie parallel to the crystal $c$-axis. The
samples were always chemically analyzed for their main
constituents. Polarized light allowed us to verify the excellent
quality of the samples and to check the orientation of the
crystallographic axes for the measurements, as determined by
$X$-ray. The sample masses in our experiments ranged from a few
milligrams to 2 g.

\begin{figure}
\begin{center}
\includegraphics[width=6cm,keepaspectratio=true]{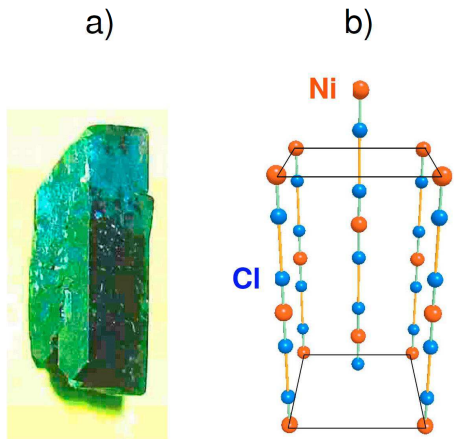}
\caption{\label{fig3}a) Sample of DTN grown in a water solution. b)
  Schematic crystal structure of the compound. The positions of
  the Ni and Cl atoms are shown.}
\end{center}
\end{figure}

Initially, to provide magnetic characterization, the magnetic
susceptibitilies of a single crystal were measured in the
directions parallel and perpendicular to the $c$-axis, and
temperatures ranging from 0.4 to 300 K.

From these data the parameters applicable to the following $S=1$
spin-Hamiltonian with single-ion anisotropy were determined:
\begin{equation}
{\cal H} = \sum_{{\bf j},\nu}J_{\nu}{\bf S}_{\bf j}
\cdot {\bf S}_{{\bf j}+e_{\nu}} + \sum_{{\bf j}}[D(S^z_{\bf j})^2
- g\mu_{B}HS^z_{\bf j}].
\label{eq:H}
\end{equation}

The obtained parameters are $g$= 2.30 and $D$ $\approx \ 9$ K.
Molecular field correction gives the exchange interaction
parameter $J$ $ \approx \ - 2.2$ K (antiferromagnetic). The high
$D$ value and the anisotropy in the exchange suggest a quasi 1D
behavior. These data were used to anticipate an induced ordered
phase at low temperature in an applied magnetic field below
$\approx \ 11$ T.

\subsection{3.1. Magnetic Induced Phase Diagram - BEC}

The field-induced phase by level crossing is permitted in
compounds that follow stringent requirements: in addition to
having a multiplet above the non-magnetic singlet, the metal ion
must reside in a site of uniaxial symmetry, so that the magnetic
field can be applied simultaneously parallel to the anisotropic
$z$ axis for all magnetic ions. The parameters for DTN and its
crystal symmetry then yield magnetic properties that can be
studied within this picture.

In DTN the high single-ion anisotropy (energy gap) is responsible
for the splitting of the $S = 1$ spin triplet of the Ni$^{2+}$ in
a $S^{z} = 0$ ground state and an $S^{z} = \pm1$ excited doublet
considerably higher in energy. The excited doublet can hop to a
neighboring doublet via the transverse component of the magnetic
exchange interaction $J$, resulting in delocalized states
analogous to the delocalized electronic states in crystals. At
zero field the small exchange interaction between the spins is not
enough to induce long-range ordering, even at zero temperature.
The application of a magnetic field $H$ leads to a splitting of
the doublet, which brings down the component parallel to the field
linearly with $H$ until the gap between the excited and ground
states closes at $H=H_{c}$. At this point magnetic excitations,
i.~e., bosons, are created (Fig. 4a).

The more realistic view of the energy levels that takes the
interaction into account is shown in Fig. 4b. The exchange
interaction produces a dispersion of the energy levels, and there
is now a field range, from $H_{c1}$ to $H_{c2}$, where the
$S^{z}=+1$ level becomes degenerate with the ground state and
bosons are created. This gapless phase extends throughout this
field range. The coupling to the magnetic field, which is
proportional to the magnetization, controls the boson density. The
BEC that appears in this intermediate field range corresponds to
the coherent superposition of the $S^{z} = +1$ doublet and the $S
= 0$ singlet at each site \cite{14,22,23,24}.

\begin{figure}
\begin{center}
\includegraphics[width=6cm,keepaspectratio=true]{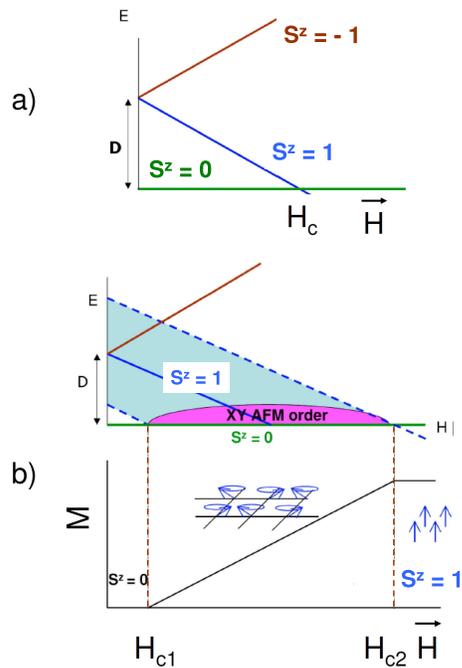}
\caption{\label{fig4} Schematic diagram ($T$=0) of the levels of
  energy. a) The lowest energy levels of $\rm{Ni^{++}}$ in an axial
  crystalline field as a function of the external magnetic field. b)
  The broad band indicates dispersion of the lowest excited level due
  to exchange coupling. Condensation occurs in the range between the
  fields $H_{c1}$ and $H_{c2}$, in which the magnetization $M$
  increases almost linearly up to the saturation.}
\end{center}
\end{figure}

The transition to the magnetic phase can be adequately described
as a condensation of bosons. Three are the basic conditions for
BEC induced by a magnetic field: 1) the excited state becoming
degenerate with the ground state, 2) strong correlation inducing
3D long-range ordering, and 3) a uniaxial symmetry leading to
number conservation of bosons.

\subsection{3.2. Hard-core Boson Model}

The first attempt to interpret a field-induced phase ordering in a
BEC was made by Nikuni \cite{7}. In his work the induced phase
diagram of $\rm TlCuCl_{3}$ was studied by applying the
Hartree-Fock-Popov (HFP) mean field analysis to the hard-core
boson Hamiltonian. For a diluted gas of particles the interatomic
interaction is sufficiently weak, therefore the mean field
Gross-Pitaevskii theory is a logical tool to study this system. In
this model the particle scattering is dominated by the two-body
contact interactions, which are described by the $s$-wave
scattering length. The physical implication of this condition is
that it is highly improbable for more particles to interact with
each other simultaneously. In hard-sphere gas, the scattering
length is equal to the diameter of the atoms. In the low-density
limit only the two particles interactions are important. The
low-energy Hamiltonian~(3) can be transformed into boson language
via identifying $S^{+}=a^{\dagger}$, where $a^{\dagger}$ is the
boson creation operator, in
\begin{equation} {\cal H} =
  \sum_{{k}}\left(\epsilon-\mu\right))a^{+}_{k}a_{k}+\frac{v_{0}}{V}\sum
  a^{+}_{k+q}a^{+}_{k´-q}a_{k}a_{k´}
\end{equation}
where $\epsilon = {\hbar k^2}/{2m}$ is the kinetic energy, $v_{0}$
is the short-range repulsion between two magnons that occupy the
same site and $V$ is the unit-cell volume of the sample.

In this hard-core model applied to the Ni$^{2+}$ ions the density
of the magnetization along the field direction, $S^{z}$ = +1, is
mapped onto the boson density. Here the $S^z=-1$ component is
neglected, because we are interested in low-field and
low-temperatures regions, where this level does not contribute. An
analysis taking this level into account is discussed in Section
4.17 \cite{14,22}.

The physics of the magnons depends on the relative strengths of
the repulsive interaction $v_{0}$ and kinetic energy. When the
interaction is attractive, $v_{0} < 0$, the system collapses. If
the repulsion interaction dominates, the system evolves into a
state where the bosons form a large lattice that gives rise to a
magnetization plateau in the $M$ $vs$. $H$ curve. If the kinetic
term dominates, the system, which corresponds to a Mott insulator
in the low-field region, undergoes a condensation at the first
critical field $H_{c1}$. Above this field the density of bosons
increases proportionally to the longitudinal component of the
magnetization $M$ up to the saturation value, $M_{sat}$, at the
second critical field $H_{c2}$. The result is an antiferromagnetic
state characterized by the presence of a staggered transverse
moment, $M_{sta}$, perpendicular to the applied magnetic field,
forming a BEC state. The absolute value and relative angle of
$M_{sta}$ are related to the amplitude and the phase of the
condensate wave function, respectively. In this phase the
condensate shows a macroscopic order parameter that spontaneously
breaks global phase symmetry by the emerging antiferromagnetic
longe-range order. In these conditions, there must exist a gapless
mode called Nambu-Goldstone in the new phase, which is necessary
to keep the original invariance of the Hamiltonian. At $H_{c2}$
all sites are occupied by bosons and the system enters a second
Mott insulating phase \cite{25,26,27,28}.

\subsection{3.3. Boson Number Conservation}

At the induced phase the Ni magnetic spins do not have any
preferred orientation in the plane perpendicular to the magnetic
field, and the antiferromagnetic order is XY-like. The tetragonal
crystal structure provides a uniaxial crystal-field symmetry about
the direction of the applied field. It is this symmetry that
enforces number conservation among the bosons. This forces the
bosons to remain in the system and macroscopically occupy the
ground state at low temperatures. Boson-number conservation is a
key condition that separates bosonic systems that condense from
those that do not. Formally, the phenomenon of BEC requires the
conservation of particle number. The boson number must be set by
some external constraint or else bosons will be excitations of the
system and vanish as the temperature is lowered to zero, as is the
case, e.g., for phonons. In the Hamiltonian (4) every creation
operator $a^{\dagger}$ is multiplied by a destruction operator
$a$. If the Hamiltonian is rotated by an angle $\phi$ in the plane
perpendicular to the field, then $a^{\dagger} \rightarrow
a^{\dagger}e^{i\phi}$ and $a \rightarrow ae^{-i\phi}$, such that
$(a^{\dagger}e^{i\phi})(ae^{-i\phi}) = a^{\dagger}a$. Because this
Hamiltonian is independent of $\phi$ the number of bosons is
conserved. Thus, the uniaxial symmetry of the Hamiltonian creates
a number conservation law for the bosons \cite{29}. In real
situations, the square lattice of the crystal can introduce a
small anisotropy in the $ab$ plane along with dipole-dipole
interactions or Dzyaloshinskii-Moriya (DM) interactions. These
effects generally occur at lower energy scales and can therefore
be neglected at the temperatures at which DTN was studied.
\cite{21} However, experimental studies should be conducted to
discard effects of this possible anisotropy in the formation of
BECs.

\subsection{3.4. Low-temperature Limit and Critical Behavior}

The self-consistent HFP approximation leads to the following
effective Hamiltonian for the magnon system:
\begin{equation}
{\cal H} = \sum_{k}
\left(\epsilon-\mu+{v_{0}}{n}\right)a^{+}_{k}a_{k},
\end{equation}
where $n$ is the density of bosons per magnetic ion, given by
$n=M/M_{sat}$.

The chemical potential
\begin{equation}
\mu=g\mu_{B}(H-H_c)
\end{equation}
determining the number of bosons in the ground state characterizes
an additional direct contribution to the magnon energy from the
external field $H$.

Bosons condense when the renormalized effective chemical potential
is zero \cite{3,7,30} (Fig. 5):
\begin{equation}
\mu_{eff}=\mu-2{v_{0}}{n_{}}=0;
\end{equation}
the factor of two comes from exchange.

\begin{figure}
\begin{center}
\includegraphics[width=6cm,keepaspectratio=true]{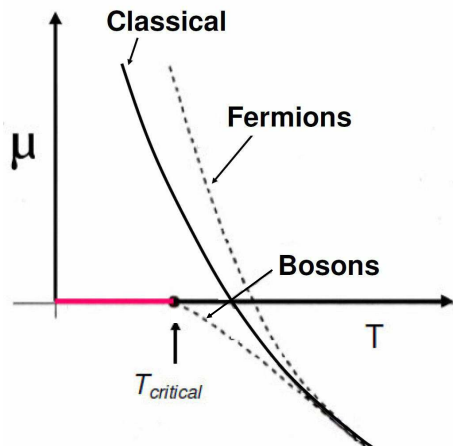}
\caption{\label{fig5} Effective chemical potential as a function of
  temperature for fermions, bosons and classical particles.
  For decreasing temperatures, $\mu$ for bosons reaches the zero limit
  at the temperature $T_\mathrm{critical}$. As $T$ becomes lower than
  $T_\mathrm{critical}$, $\mu$ is pinned at zero \cite{31}}
\end{center}
\end{figure}

At this "point of condensation", where the gap is closed, the
applied magnetic field is responsible for driving the system to
the BEC phase in which bosons are created at ($H_{c1},T_{c1}$). At
$H_{c1}$ the magnetization acquires the value $M_{c}(T)$ and
$n_{c}(T)= M_{c}(T)/M_{sat}$ gives the number of bosons created at
the transition. Using (6) and (7), the low-field boundary of the
phase diagram $H$ $vs$.~$T$ can be written as \cite{3,20,21,25,26}
\begin{equation}
H_{c}(T)-H_{c}(0)= [2v_{0}/g\mu_{B}]n_{c}(T).
\end{equation}

On the other hand, the density of created bosons is given by
Bose-Einstein statistics
\begin{equation}
  \rho(T)= (1/V) \sum_{{\bf k}}[1/[exp(\epsilon_{k}/T)-1)]
\end{equation}

That is, at the low-temperature limit \cite{1,7,9,32} and letting
the chemical potential be $\mu_{eff}$:
\begin{equation}
\rho(T)\approx
\zeta(3/2)\left(\frac{mk_{B}T}{2\pi\hbar^{2}}\right)^{3/2},
\end{equation}
since the density of bosons per magnetic ion is $n_{c}(T)$=
$\rho(T)V$, where $V$ is the unit cell volume,~(8) becomes
\begin{equation}
H_{c}(T)-H_{c}(0)\approx [2v_{0}/g\mu_{B}]
\zeta(3/2)\left(\frac{mk_{B}T}{2\pi\hbar^{2}}\right)^{3/2}V.
\end{equation}

\subsection{3.5. Signature of the BEC}

Based on this analysis, the phase boundary can be interpreted as a
BEC of magnons. At low temperatures only the lowest excited energy
level is considered, so that the interaction can be replaced by
the constant $v_{0}$. In this case, introducing a hard-core
constraint assuming an infinite on-site repulsion, only one boson
occupies each site.  Certain properties of the induced magnetic
phase in $\rm TlCuCl_{3}$ can be explained by the concept of a
BEC. Plotted as a function of the field, the magnetization shows
the cusplike dip at the transition predicted by HFP theory
\cite{7,20,26}. However, disagreements are found with other
experiments, wich were later attributed to the lack of uniaxial
symmetry.

One of the few compounds entering the BEC phase under magnetic
fields sufficiently low to be reached using standard
superconducting magnets is DTN. Most other candidates require
pulsed fields and yield much less accurate experimental results.
To condensate at $T\neq 0$, magnons should be able to propagate in
three dimensions. For BEC to occur in quasi-1$D$ system, such as
DTN, there must be weak interaction between the 1$D$ chains, which
permits full mobility of magnons in 3$D$.

One of the experimental features of the BEC of magnons can be
found by analyzing the experimental data at the boundary
$H_{c1}(T)$, where part of the excitations condense.

The signature of the BEC universality class at low temperature is
given by the power-law \cite{3,7,12,20,21,25,26}
\begin{equation}
H_{c1}(T)-H_{c1}(0) \approx \ A T^{\phi},
\end{equation}
where $A$ is fully determined by the coefficient of (11) as a
function of $v_{0}$ and the mass of magnons.

DTN behaves in 3D below $1.2$ K as a BEC, which is where the
long-range order sets in, although the universal behavior with
$\phi = 3/2$ is observed only below $T < 0.2$ K. This is because
the $\phi = 3/2$ power-law behavior is a low-temperature
approximation to the full boson distribution function.

\section{IV. EXPERIMENTAL DETERMINATION OF THE BEC PHASE}

A wealth of useful information about the nature of the transition
to the ordered phase can be obtained from several thermodynamics
properties in the temperature $vs$. field plane (Fig. 2). The
first thermodynamic property used to study the transitions is the
magnetization, one of the most fundamental quantities observed in
experiments, followed by specific heat and magnetocaloric effect.
More properties such as magnetostriction, thermal transport, EPR,
inelastic neutron scattering, and other techniques are used to
identify the induced magnetic phase as BEC.

\subsection{4.1. Magnetization}
The boundary of the induced phase diagram of DTN can be
experimentally determined by curves of magnetization $vs$.
temperature at fixed $H$, with a cusp-like minimum at $T_{c}$, or
by $H_{c}$ at the inflexion in scans of magnetization at fixed
temperature $T$. These critical points originate from the
competition of two behaviors of the magnetization, which
correspond to the number of magnons: above $T_{c}$, magnetization
is determined by the number of thermally excited magnons, which
decreases with decreasing $T$; below $T_{c}$ the number of
condensed magnons increases with decreasing temperature,
overcoming the decrease of thermally excited magnons. As a result
of this competition on the origin of the magnons, $M$ has a
minimum at $T_{c}$, with the value $M_{c}$ (Fig. 6a). For the same
reason, when the phase boundary is crossed by sweeping $H$ at a
fixed temperature, the condensation occurs at the inflexion field
$H_{c}$ (Fig. 6b). It has been proved that at $H_{c}(T_{c})$ a
discontinuity of the magnetization is provided by the sudden
vanishing of the chemical potential $\mu$ \cite{7,20,25,26,28}.

The magnetization was measured using a vibrating sample
magnetometer, VSM, in temperatures above 0.4 K  at the High
Magnetic Field Laboratory facility of the University of S\~{a}o
Paulo. The sample was immersed in a $^{3}$He bath, and the
temperature was measured by vapor pressure and carbon-glass or
cernox thermometer. The magnetic field was supplied by a
Nb$_{3}$Sn superconductor. Measurement at $T = 0.019$ K was made
using the gradient field method in the dilution refrigerator. The
transitions were determined in the magnetization versus field
curves by locating anomalies in the second derivative of the
magnetization \cite{33}.

\begin{figure}
\begin{center}
\includegraphics[width=7.5cm,keepaspectratio=true]{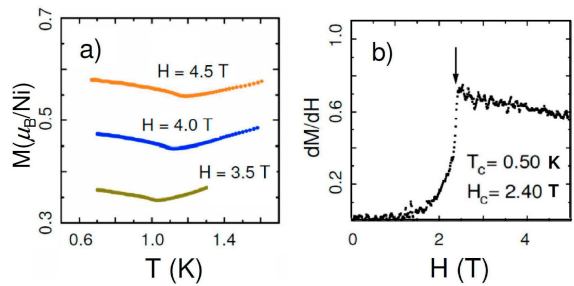}
\caption{\label{fig6} a) Curves of magnetization taken at several
  fields. The minimum indicates the transition temperature $T_c$. b)
  Representative curve of $dM/dH$. The sharp bending indicates the
  transition field $H_c$.}
\end{center}
\end{figure}

\subsection{4.2. Specific Heat}
All measurements were made on crystals with the external field $H$
provided by the 17-20 T superconducting magnet in a dilution
refrigerator system at the National High Magnetic Field Laboratory
in Los Alamos. The magnetic field was always applied along the
tetragonal $c$-axis of the sample. Specific heat was measured with
two methods, as shown in Fig. 7.

Measurements at quasi-constant temperature were made using a
developed ac technique, Fig. 7a. A small sample, $\approx1$ mg,
was mounted onto a home-made calorimeter consisting of a Si
platform furbished with an amorphous-metal thin film heater and a
bare Cernox thermometer. The calorimeter was attached to the
temperature regulation block and was held in a cryogenic vacuum
system inside a refrigerator. The sample heater is driven by one
lock-in amplifier running at frequency $f$, and the temperature is
detected with a second lock-in locked to the frequency of 2$f$. In
both cases a peak is observed at the transitions.  The observed
anomalies in $C_{p}$ map the known phase boundaries in DTN quite
well. In the $C_{p}$ versus $H$ curves a double peak is apparent,
with the peak near $H_{c2}$ far larger than the peak near
$H_{c1}$. A comprehensive theory of the specific heat has been
developed. Quantum fluctuations present near $H_{c1}$, but not
near $H_{c2}$, reduce the kinetic energy of the magnons, causing a
strong mass renormalization near the two transitions,
$m(H_{2})/m(H_{1})\approx 3$. This effect, closely described by
analytical and quantum Monte Carlo calculations, explains similar
asymmetries observed in other properties of DTN, such as
magnetization, electron spin resonance, magnetostriction, and
thermal conductivity \cite{34}.

At constant field the method used was the quasiadiabatic heat
pulse relaxation technique (Fig 7b). The sample was mounted onto a
sapphire plate and its temperature was monitored by a RuO$_{2}$
field-calibrated thermometer \cite{14}.

\begin{figure}
\begin{center}
\includegraphics[width=7.5cm,keepaspectratio=true]{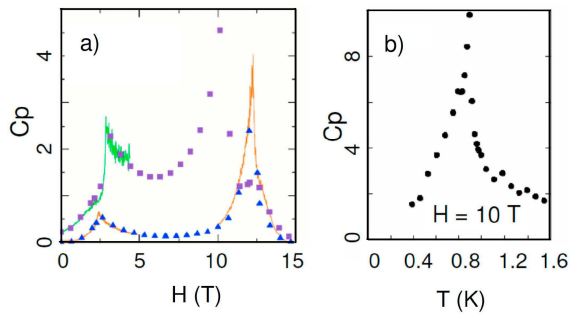}
\caption{\label{fig7} a) Specific heat as a function of the magnetic
  field for fixed temperatures of 0.40 K ( triangles) and 0.75 K (
  squares).  b) Specific heat in an applied field of 10 T.}
\end{center}
\end{figure}

\subsection{4.3. Magnetocaloric Effect }

This effect was measured by sweeping the field up and down while
monitoring the sample temperature with the bath temperature held
fixed. The device was the same that was described in measurements
of $C_{p}$ by the relaxation technique \cite{14}. In the
magnetocaloric effect, whose data we show in Fig.~8a, heating is
observed as the magnetic field is swept through the boundary of
the phase diagram, surrounded by regions of cooling before and
after the transition. The peak in the first derivative of $T(H)$,
Fig. 8b, corresponding to the maximum heating of the sample, was
identified as the phase transition.

\begin{figure}
\begin{center}
\includegraphics[width=7.5cm,keepaspectratio=true]{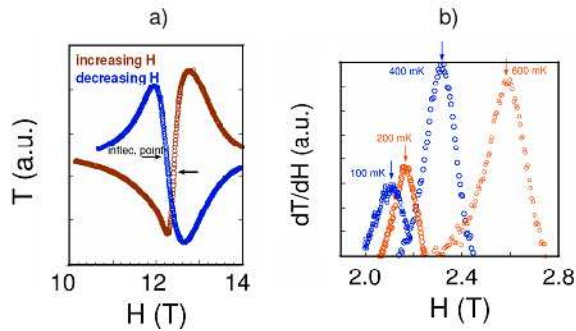}
\caption{\label{fig8} a) Magnetocaloric effect determined by
  monitoring $T$ while sweeping $H$ up and down. b) derivative $dT/dH$
  for several temperatures, where the transition is identified as the
  peak.}
\end{center}
\end{figure}

\subsection{4.4. Phase Diagram }

The induced ordered phase in the field range $H_{c1} < H < H_{c2}$
extends to a finite temperature $T_c(H)$ in which the maximum
temperature for DTN is $T_{max} = 1.2$ K. The thermal and quantum
phase transitions are qualitatively different.  In the
field-induced quantum critical point, the magnetic order is
suppressed by both the phase and the amplitude fluctuations.
Figure 9 shows the constructed phase diagram.

The measured thermodynamic properties of DTN $C_{p} (H, T)$, $M
(H, T)$, $\Delta T(H)$, used to characterize the transitions, were
analyzed in regions near the phase boundaries, the critical
regions, to confirm that this is a phase transition to a BEC. Very
few quantum magnet candidates to BEC have been studied at
temperatures in which the power-law dependence can be confirmed.
If the temperature range in which the power-law fit is performed
is far from zero temperature, it is difficult to accurately
identify the power-law behavior with correct parameters. For
example, the first determination of the exponent in DTN yielded
$\phi\approx 2$, very different from the expected value \cite{24}.
This problem can be circumvented by the windowing method, in which
the intercept and the exponent are determined independently by
performing fits over different temperature ranges and
extrapolating the values to zero temperature. The results show
that the field-temperature phase boundary approaches a power law
near the quantum critical point, with an exponent value
$\phi\approx 1.5$, as expected for 3$D$ BEC \cite{14}.

Using the theoretical prediction obtained for the scaling form of
the field- and temperature-dependent magnetization close to
$H_{c2} (T)$, an analysis of the magnetization shows a good
agreement between the theoretical and experimental results. This
clearly shows that the transition at $H_{c2} (T)$ is a BEC of
magnons \cite{35}.

\begin{figure}
\begin{center}
\includegraphics[width=7cm,keepaspectratio=true]{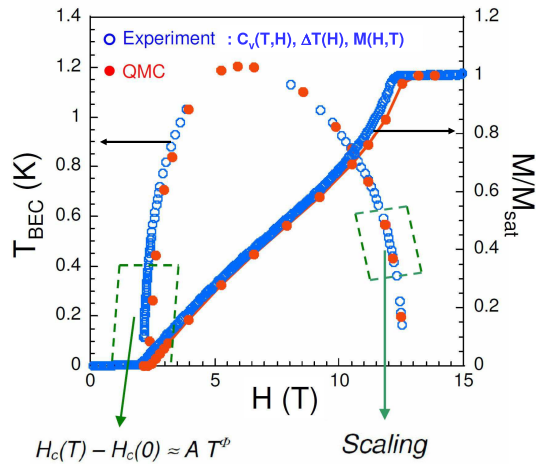}
\caption{\label{fig9} Magnetic phase diagram determined from specific
  heat, magnetocaloric effect, and quantum Monte Carlo simulations
  (QMC). The magnetization $vs.$~field measured at $19$ mK
  and calculated from QMC simulations are overlaid on the phase
  diagram \cite{14}.}
\end{center}
\end{figure}

\subsection{4.5. Neutron Scattering}

A microscopic understanding of the ground state was reached
through inelastic neutron measurements. Low-energy magnetic
excitations were performed using the cold neutron triple-axis
spectrometer at the Paul Scherrer Institute, Switzerland. Using an
Oxford instruments dilution insert in a VARIOX cryostat, a sample
of three co-aligned deuterated single crystals of DTN with a
combined mass of 3 g were cooled to $T= 80$ mK. The analysis of
the resulting dispersion of the magnetic excitation was adjusted
to the Hamiltonian with interaction. The observed scattering was
analyzed using a single-mode cross section with dispersion
obtained from a generalized spin-wave approach for the disordered
phase. The obtained parameters agree with a model of spins of Ni
ions strongly coupled along the tetragonal axis, but only weakly
perpendicular to it, making DTN a weakly coupled chain system with
an anisotropy higher than the exchange interactions \cite{14}.

\subsection{4.6. Electron Spin Resonance}

We availed ourselves of the high-field approach exact theoretical
expressions for the spin-polarized phase. Investigating magnon
excitations in fields up to 25 T allowed us to obtain a reliable
set of spin-Hamiltonian parameters. Part of this work was
performed at the National High Magnetic Field Laboratory,
Tallahassee. The parameters agree very well with those obtained
from fitting the experimentally induced phase boundary and
low-temperature magnetization of DTN with results of quantum Monte
Carlo simulation. The parameters were used to calculate the
frequency-field dependence of two magnon bound-state excitations
predicted by theory, and observed in DTN for the first time
\cite{36,37,38}.

\subsection{4.7. Four Sublattice Model}

We present the low-energy excitation spectrum in the magnetically
ordered phase obtained from spin resonance measurements down to
0.45 K. The EPR measurements were made at the Kapitza Institute
using a transmission-type spectrometer equipped with a cylindrical
multimode resonator and a $^{3}$He cryostat. The observed modes
can be interpreted within a four-sublattice antiferromagnetic
model with a finite interaction between two tetragonal subsystems,
$V_\mathrm{intra-sub}$ \cite{36}.

\subsection{4.8. Crossover from the 1D Fermionic to the 3D Bosonic
  Character in DTN}

Magnetoacoustic studies in the vicinity of the quantum critical
points in DTN show that the behavior of the observed properties
outside the ordered phase boundary can be well described by an
effective 1D fermionic model of low-lying spin excitons. This,
together with previous results showing the bosonic behavior inside
the boundary, suggests a crossover from the bosonic to the
fermionic character of the magnetic excitations close to the
quantum critical points. One fascinating aspect of this crossover
is continuous tuning from the bosonic to the fermionic statistics.
This research was conducted in Dresden \cite{39}.

\subsection{4.9. Critical Behavior at the Low Field Transition}

Interpretating the low-field phase boundary as a transition to the
BEC phase allows one to obtain information about the boson
interaction in DTN.

The critical field $H_{c}$, plotted as a function of $n_{c}$ in
Fig.~10a, follows the linear relation in~(8). The interaction
strength constant $v_{0}$ in that equation can be obtained from
the derivative of the fitted curve $dH_{c}/dn_{c} = 9.52 =
2v_{0}/g\mu_B= B$ as $v_{0}$ = 0.61 meV. The zero temperature
transition field is obtained as $H_{c}(0)= 2.15$ T, in close
agreement with Ref.~[14]. This confirms the behavior is linear
over the entire temperature range. Using the value of $A \approx
0.70 (T/K^{3/2})$, as obtained from the phase diagram at low
temperature, and $B$ = 9.52, ~(8)~and (12) simultaneously yield
the derivative $dn_{c}/dT^{3/2} =A/B = 0.073$, which is plotted in
Fig.~10b as a straight line. The continuity of this line at low
temperature, with the experimental points at high temperature, is
a robust indication that the theory is sufficiently coherent to
allow us to combine results from different sets of experimental
data \cite{21,40}.

\begin{figure}
\begin{center}
\includegraphics[width=7cm,keepaspectratio=true]{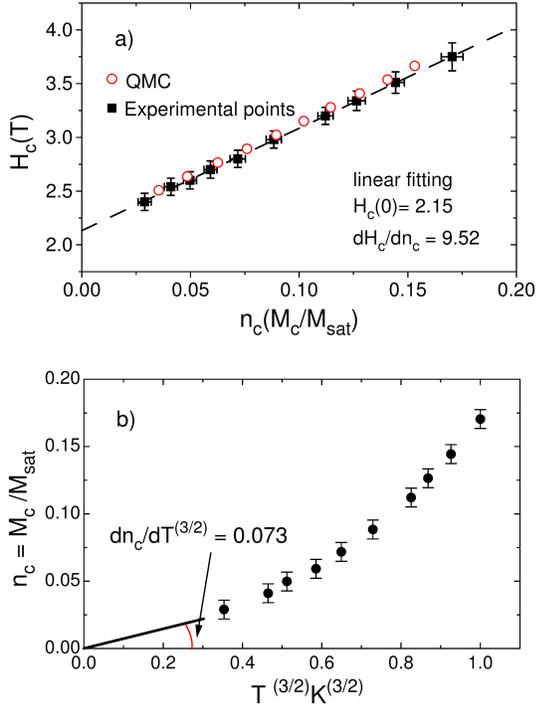}
\caption{\label{fig10} a) Critical transition field as a function of
  critical magnetization density $n_{c}$. The $filled$ $squares$ represent
  the measurments. The $dashed$ $line$ denotes the linear fitting of the
  data. The Quantum Monte Carlo data are shown as $red$ $circles$. b)
  $n_{c}$ as a function of $T^{3/2}$. The $straight$ $line$ is
  the plot of $dn_{c}/dT^{3/2} =A/B = 0.073$, obtained from (8) and (12) from different experiments. The continuity of this
  derivative with the experimental points is very high \cite{40}}.
\end{center}
\end{figure}

\subsection{4.10. Boson Number Conservation}
The derivative $dM/dH$ in Fig.~6b shows a sharp kink at $H_{c}$.
In a recent report, a Dzyaloshinskii-Moriya (DM) interaction in
DTN was proposed to explain the extra lines in ESR experiments.
This anisotropic interaction between corner-center coupling spins
at sites $i, j$ in the body-centered tetragonal lattice of DTN is
given by $\sim$
\textit{$\textbf{d}\cdot(\textbf{S}_{i}\times\textbf{S}_{j})$},
where \textsl{$\textbf{d}$} is a specific vector coefficient. For
any possible DM interaction, the vector \textsl{$\textbf{d}$}
would have to point along the tetragonal axis. Otherwise a
nonuniform field distribution would exist in the sample, which
would induce a staggered moment perpendicular to the applied
field, which in turn would broaden the magnetization at the
transition. This configuration characterizes a broken axial
symmetry, as is observed in $\rm TlCuCl_{3}$. Traces of $M$ $vs$.
$H$ are the best for analyzing the configuration of the vector
\textsl{$\textbf{d}$} in DTN. From the sharp peak in
$d^{2}M/dH^{2}$ at the transition, we conclude that any possible
$U (1)$ symmetry-breaking terms are small at these temperatures,
and the number of bosons is conserved, as required for BEC
\cite{21}.

\subsection{4.11. Magnetostriction}

Magnetostriction taken at dilution refrigerator temperatures,
performed in a titanium dilatometer, shows significant
magnetoelastic coupling and a magneto-order-induced modification
of the lattice parameters in DTN. Length changes were monitored
capacitatively with a Be-Cu spring-mounted titanium tip in a
plastic rotator. The top-load dilution refrigerator was mounted in
a 20 T superconductor magnet system at the National High Magnetic
Field Laboratory in Tallahasse. A simple theory relates the
magnetostriction to the spin-spin correlation function and
describes the data remarkably well. From this, was possible
extract the spatial dependence of the magnetic exchange coupling.
The measured data agree excellently with those obtained from
quantum Monte-Carlo simulations and with the phase diagram
determined from $C_{p}$ and magnetostriction \cite{41}.

\subsection{4.12. Direct Measurement of the BEC Universality Class}

Although the windowing method yields the critical exponent at zero
temperature by an efficient extrapolation procedure, the physics
that can appear below the studied temperatures can be hidden by
this method. This was clearly demonstrated for $\rm
BaCuSi_{2}O_{6}$, in which a reduction in dimensionality of the
spin system changed the exponent, as was observed at the lowest
temperature. In DTN new measurements were performed at
temperatures down to 1 mK, which is two orders of magnitude below
the lowest temperature scale for magnetic coupling in this system,
$J_{ab} = 0.18$ K, and the lowest ever used to investigate BEC in
a quantum magnet. Thus, with this new experiment, no extrapolation
was needed to determine the power-law exponent. Measurements of
the transition were carried out using ac susceptibility. The
experiments were performed using a $\rm PrNi_{5}$ nuclear
refrigerator and a 15 T magnet at the National High Magnetic Field
Laboratory Facility at the University of Florida. The sample was
immersed in liquid $^{3}$He in a polycarbonate cell, rather than
being glued to a cold finger, and thermal contact to the
refrigerator was assured via sintered silver. The temperature was
calibrated using a $^{3}$He melting-pressure curve thermometer.
The device used for operation at ultra-low temperatures is the
mutual inductance bridge assembly. The transitions were determined
by the derivative of the susceptibility curve. The power-law
temperature dependence of the phase transition line, with $\phi =
3/2$, was found to be consistent with the 3$D$ BEC universality
class, which confirms the conclusion obtained by the windowing
method \cite{42}.

\subsection{4.13. Disorder in the Doped System DTN/Br}

When disorder is introduced in a magnet with interacting bosons,
the condensate disrupts and interferes with phase coherence. The
result is the creation of a peculiar state, the Bose-glass phase
(BG), with only short-range correlation. In the presence of
disorder the transition to BEC is argued to occur only from the
BG, and never from the Mott insulator. The transitions to BEC
should be governed by physical parameters characteristic for BG
$\rightarrow$ BEC for the low-field transition and for BEC
$\rightarrow$ BG for the high-field transition. Experimental and
theoretical studies were realized in the disorder quantum magnet
DTN/Br, Ni(Cl$_{1-x}$Br$_{x}$)$_{2}$-4SC(NH$_{2}$)$_{2}$. The
remarkable agreement between theory and experiment shows a
well-controlled realization of a disordered Bose fluid with a new
seemingly universal exponent governing the scaling of the critical
temperature from BG to BEC \cite{43,44}.

\subsection{4.14. Mass of Magnons}

From the linear behavior of $n_{c}$ at low temperatures in Fig.
10b, the effective mass of the bosons in DTN, $m_{b}$, can be
calculated with the coefficient $dn_{c}/dT^{3/2}$ from (8) and
(11):

\begin{equation}
  dn_{c}/dT^{3/2}=\zeta(3/2)\left(\frac{m_{b}k_{B}}{2\pi\hbar^{2}}\right)^{3/2} V = 0.073.
\end{equation}

The obtained value is $m_{b}$=5.4x$10^{-25}$ g, which is
approximately 1/3 of the proton mass. The value for $m_{b}$ can be
compared with that obtained from neutron dispersion data,
$m$=3.7x$10^{-25}$ g \cite{45,46}.

\subsection{4.15. Quantum Depletion of Magnons}

Although the repulsive interaction between bosons are essential to
avoid the collapse of particles, $v_{0}$ depletes the number of
condensed bosons by the quantity of non-condensed bosons. This was
described by Nikuni, who obtained the density of bosons in the BEC
phase from the sum of the condensed and the non-condensed
densities.  To determine the fraction of condensed bosons, we
analyzed the $n$ $vs$.~$H$ curve at $T=0.019$ K in Fig.~11, which
is reasonable approximation to the $T\rightarrow 0$ limit. The
relation between the total density of particles $n$ and the
density of condensed bosons $n_{cond}$, as an effect of $v_{0}$,
is given at zero temperature by the equality
\begin{equation}
  n(0)=n_{cond}(0)+ \frac{1}{3\pi^{2}}
  \left(\frac{mv_{0}n_{cond}(0)}{\hbar^{2}}\right)^{3/2}
\end{equation}.

The second term on the right-hand side of this equation represents
the number of condensed bosons that are scattered out of the
ground state due to interaction $v_{0}$. The non-condensed
fraction at $T=0$, called quantum depletion, $\delta n$, is caused
by quantum fluctuations around the true condensate. Applying~(14)
to the experimental data we determined the dependence of the
condensed bosons with the magnetic field.  The open circles in
Fig.~11 show the calculated density of condensed bosons. The
depletion of condensed bosons just above the critical field
$H_{c}(0)= 2.09$ T increases almost linearly up to $\approx$ 10
per cent at the field of $3.0$ T \cite{45}.

Depletion is an important matter in BEC because the dissipation
effects of non-condensed bosons leads to the creation of solitons
and vortices in diluted BECs. In DTN, an insulating system, this
effect has not been observed.

\begin{figure}
\begin{center}
  \includegraphics[width=7cm,keepaspectratio=true]{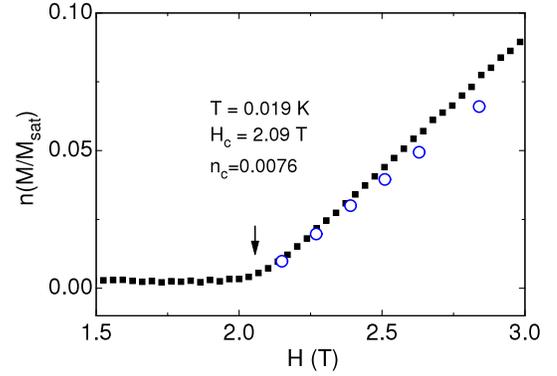}
  \caption{\label{fig11} Magnetization density at $T$= 0.019 K as a
    funtion of the field. The $filled$ $squares$ represent the measured
    densities. The $open$ $symbols$, representing the condensed bosons,
    are obtained from~(14)}
\end{center}
\end{figure}

\subsection{4.16. Theoretical Methods}

Almost throughout, the measurements are compared with quantum
Monte Carlo data. The simulations agree exceptionally well with
the experiments. Some measurements also are correctly fitted by
analytic expressions. The values for the DTN parameters refined
via QMC and ESR are $D= 8.9$ K, $J_{c}= 2.2$ K and $J_{ab} = 0.18$
K \cite{38}.

With an alternative approach using the diagram technique, one
finds an expression for the magnetic phase boundary. Taking into
account the coupling constant between the two different tetragonal
sublattices, the following set of parameters fit the experimental
data very well: $D= 7.72$ K, $J_{c}= 1.86$ K, $J_{ab} = 0.2$ K,
and $V_\mathrm{intra-sub} \approx 0.1$ K \cite{47}.

\subsection{4.17. Semi-hard-core Boson Model}

Both $\rm TlCuCl_{3}$ and $\rm BaCuSi_{2}O_{6}$ have a strong
antiferromagnetic coupling between the two $S$=1/2 spins in each
Cu dimer, forming an $S$=0 singlet ground state separated by a gap
$J$ from the $S$=1 excited triplet. The hard-core model describes
the experimental symmetry-induced phase diagram of these systems
very well. In DTN this model describes the behavior of the
transitions only at low temperatures, $T < 0.2$ K, and low fields,
$H\alt H_{c1}$. An alternative approach can be used to describe
the asymmetry of the phase diagram at high fields mapping the
$S=1$ system onto a gas of semi-hard-core bosons. This asymmetry
is attributed to the influence of the highest energy level and is
not expected for the effective model with $S$=1/2. In this model
the maximum boson population is limited to two per lattice site.
More studies focus on the analysis of the upper critical field and
a self-consistent theory \cite{48,49}.

\subsection{4.18. Two-magnon Bound States}

Theoretical and experimental systematic high-field ESR studies of
DTN have been presented with special emphasis on single-ion
two-magnon bound states. To clarify some remaining discrepancies
between theory and experiment, we analyzed the frequency-field
dependence of the magnetic excitations in this material. In
particular, a more comprehensive interpretation of the
experimental signature of single-ion two-magnon bound states is
shown to be fully consistent with theoretical results. Moreover,
we have clarified the structure of the ESR spectrum in the
so-called intermediate phase. The experimental part was performed
at the National High Magnetic Field Laboratory, Tallahassee.
\cite{50}

\section{V. SUMMARY}

We presented an experimental review of the recently observed
behavior of DTN associated with BEC phenomena. These experiments
provide opportunities for an introductory study of the BEC of
magnons in insulating compounds when bosons are induced by an
applied magnetic field. The simple magnetic structure of DTN
enables experiments in field and temperature regions that are
easily accessible in the laboratory. As shown in the theoretical
outline presented here, the many-body aspects of BEC are reduced
to an effective single-particle description, where interactions
give rise to an additional potential proportional to the local
particle density. This model allows characterizing the
condensation of magnons at low temperatures through parameters
such as critical exponents, analysis of the boson density, boson
interaction strength, boson mass, depletion, etc. Although this
hard-core model describes the low-field behavior of DTN very well,
the whole phase diagram can be exactly explained by quantum Monte
Carlo simulations with a semi-hard-core model.

The purpose of this review was to describe experiments that go
beyond the mean-field theory and explore effects that modify the
transition to the field-induced magnetic phase.

Finally, we note that the research in this field has not been
limited to condensed matter, but was extended to the applications
in nuclear physics, astrophysics, and particle physics.

\section{ACKNOWLEDGEMENT}

The success of the realizing so many experiments on DTN is the
result of a collective effort of expertise in the field of
magnetism and low temperature. I especially appreciate the
stimulating discussions with Vivien Zapf, Marcelo Jaime, Liang
Yin, and Tommaso Roscilde throughout the friendly collaboration on
the experimental research of BEC. Many other researchers were
involved in the studies listed here, including important support
by theoretical works.

Investigating the Bose-Einstein condensation in quantum magnets is
part of the author's research program, which is supported by
FAPESP and CNPq at the Solid-State and Low-Temperature Laboratory
of the University of S\~{a}o Paulo. Part of this work was
performed at the National High Magnetic Field Laboratory, which is
supported by National Science Foundation Cooperative Agreement No.
DMR-0654118, the State of Florida, and the U.S. Department of
Energy.


\begin{thebibliography}{10}
\bibitem{46} The boson mass was calculated by M.~Kenzelmann from the
  neutron-scattering data in Ref.~14. Since at low
  temperature only the bottom of the magnon band is thermally excited
  around one minimum, $m$ can be evaluated from the curvature of the
  band structure around this point. The dispersion is given by $E =
  \Delta + c^{2}k^{2}$, where $\Delta$ is the excitation gap and $c$
  describes the increase in the dispersion as a function of the
  wave-vector $k$ in lattice units. Writing $E_{c}=p^{2}/(2m)$, the
  obtained mass is $m=(1/2)({2\pi\hbar}/(a/c))^{2}=3.7\times10^{-25}$ g.
\end{thebibliography}
\end{document}